\documentclass[a4paper]{article}

\usepackage[english]{babel}
\usepackage[utf8x]{inputenc}
\usepackage[T1]{fontenc}
\usepackage[affil-it]{authblk}
\usepackage{hyperref}

\usepackage{mathtools}

\DeclarePairedDelimiter\floor{\lfloor}{\rfloor}

\usepackage[a4paper,top=3cm,bottom=2cm,left=3cm,right=3cm,marginparwidth=1.75cm]{geometry}

\usepackage{amsmath}
\usepackage{graphicx}
\usepackage[colorinlistoftodos]{todonotes}

\title{SweetRS: Dataset for a recommender systems of sweets}
\author{{\L}ukasz Kidzi\'nski\thanks{Electronic address: \texttt{lukasz.kidzinski@stanford.edu}}}
\affil{School of Engineering, Stanford University, USA}

\begin{document}
\maketitle

\begin{abstract}
Benchmarking recommender system and matrix completion algorithms could be greatly simplified if the entire matrix was known. We built a \url{sweetrs.org} platform with $77$ candies and sweets to rank. Over $2000$ users submitted over $44000$ grades resulting in a matrix with $28\%$ coverage. In this report, we give the full description of the environment and we benchmark the \textsc{Soft-Impute} algorithm on the dataset.
\end{abstract}


\section{Context}

One of the problems in building any machine learning system is the limited access to ground truth. This problem is particularly prevalent in matrix completion when the matrices are very sparse, such as in the context of product recommendation. In many situations, a large dataset can be trimmed to a dense matrix by choosing specific users and items, yet it may introduce additional bias. In this project, we attempted to collect a dense matrix, by rating commercial products internationally known, such as candy bars and sweets.

We built a basic website \url{sweetrs.org}, where users can both rate and add new products. Participants were tasked to rate sweets on the scale from $1$ to $5$ or click "Never tried" in case they do not know or have not tasted the product. To the date, we collected over $44000$ ratings from over $2000$ users on $77$ items, giving the coverage of over $28\%$ matrix coefficients. Moreover, we identified a subset of $677$ users and $45$ products with the coverage of over $95\%$ matrix coefficients.

The project has been developed as a part of a Master Thesis at the University of Warsaw \cite{kidzinski2011statistical}.

\section{Benchmark}

Let $X_{N \times D}$ be a matrix representing $N$ users rating $D$ items on the $1-5$ integer scale. Let $\Omega$ be a set of all observed indices $(i,j)$. We attempt to approximate unobserved ratings. We estimate prediction error using cross-validation.

As our benchmark method we choose \textsc{Soft-Impute} \cite{mazumder2010spectral} due to its speed, efficiency and simplicity. \textsc{Soft-Impute} performs thresholded SVD in the presence of missing values. 

We investigate how the prediction depends on the size of the training set and on the regularization parameter $\lambda$ in \textsc{Soft-Impute}. We test $9$ settings for the size of the training set, with $p \in \mathcal{P} = (0.1,0.2,...,0.9)$ where $p$ is the ratio of the observed set used for testing. 

We use cross-validation for estimating the Normalized Mean Squared Error:
\begin{align*}
NMSE(\hat{X}) = \sum_{(i,j) \in \Omega - \Gamma} \| \hat{X}_{i,j} - X_{i,j} \|^2 / \sum_{(i,j) \in \Omega - \Gamma} \| X_{i,j} \|^2,
\end{align*}
where $\Gamma$ is the set on which we trained the algorithm and $\Omega - \Gamma$ denotes the set difference.

\begin{figure}[h]
\centering
\includegraphics[width=0.9\textwidth]{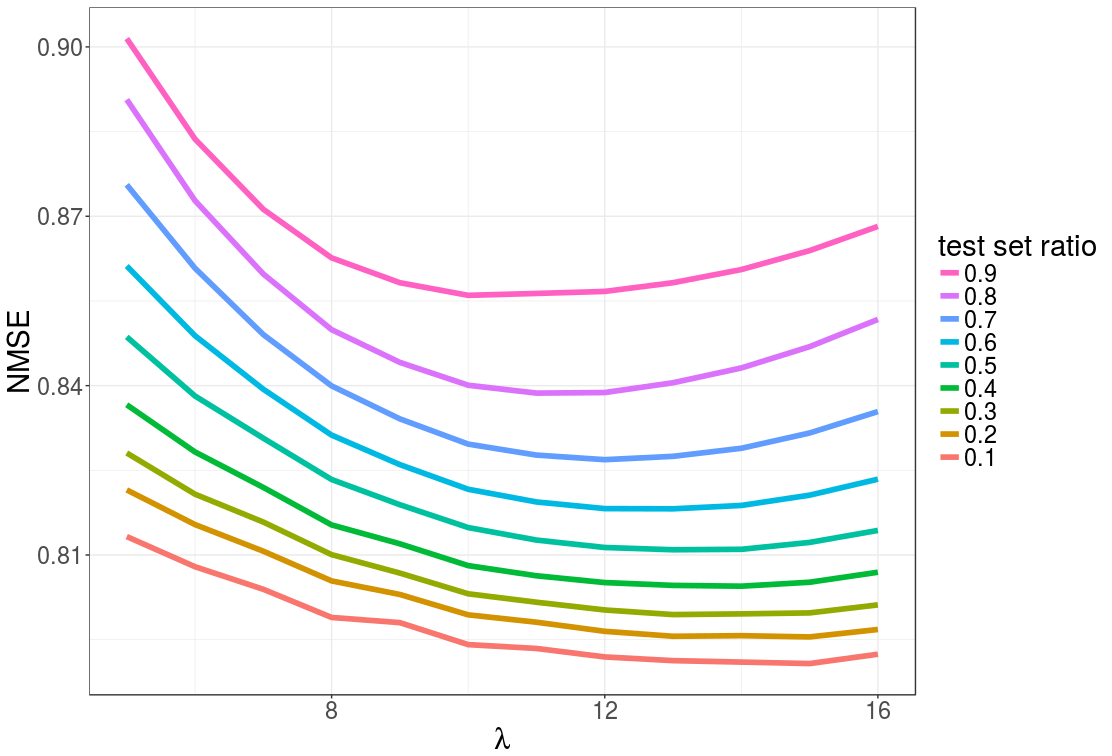}
\caption{NMSE as a function of $\lambda$ for different sizes of the training set. Values were estimated by cross-validation repeated $10$ times for each combination of parameters.}\label{fig:nmse}
\end{figure}

In our preliminary experiments, we identified that analyzing the set $\Lambda = (1,2,...,20)$ is sufficient for finding best integer $\lambda$. Thus, we train the \textsc{Soft-Impute} algorithm on constellations $\mathcal{P} \times \Lambda$. We center and scale each item before fitting. Next, we estimate NMSE for a given $(p,\lambda)$ is performed as follows:
\begin{enumerate}
\item randomly choose $\floor{p\cdot|\Omega|}$ training elements $\Gamma_p$,
\item fit the \textsc{Soft-Impute} model for on $\Gamma_p$ given the parameter $\lambda$,
\item predict elements $\hat{X}_{i,j}$ for $(i,j) \in \Omega - \Gamma$,
\item record $NMSE(\hat{X})$.
\end{enumerate}
We repeat the procedure $10$ times for every $(p,\lambda) \in \mathcal{P} \times \Lambda$. We present mean $NMSE$ in Figure \ref{fig:nmse}.

\section{Discussion}

In this report we aimed at providing a benchmark and description of the dataset convenient for testing new recommender system techniques. We achieved over $20\%$ of variance explained in the case when $90\%$ of matrix coefficients are observed. We published the dataset and the sample \verb|R| code as a github repository\footnote{\url{https://github.com/kidzik/sweetrs-analysis/}}.

\bibliographystyle{alpha}

\end{document}